# Initiating Heavy-atom Based Phasing by Multi-Dimensional Molecular Replacement

*Bjørn Panyella Pedersen[*], Pontus Gourdon, Xiangyu Liu, Jesper Lykkegaard Karlsen &
Poul Nissen*

*Centre for Membrane Pumps in Cells and Disease, Danish National Research Foundation,
Dept. of Molecular Biology, Aarhus University.
Gustav Wieds Vej 10C, DK – 8000 Aarhus, Denmark
* Correspondence e-mail: bpp@mb.au.dk.*

### Synopsis

A strategy is presented to to set up an n-dimensional Molecular Replacement Parameter Matrix (MRPM) search, using objective signals to uncover weak, but correct, molecular replacement solutions that can be used for heavy atom site identification and subsequent experimental phasing.

### Abstract

To obtain an electron-density map from a macromolecular crystal the phase-problem needs to be solved, which often involves the use of heavy-atom derivative crystals and concomitantly the determination of the heavy atom substructure. This is customarily done by direct methods or Patterson-based approaches, which however may fail when only poorly diffracting derivative crystals are available, as often the case for e.g. membrane proteins. Here we present an approach for heavy atom site identification based on a Molecular Replacement Parameter Matrix (MRPM) search. It involves an n-dimensional search to test a wide spectrum of molecular replacement parameters, such as clusters of different conformations. The result is scored by the ability to identify heavy-atom positions, from anomalous difference Fourier maps, that allow meaningful phases to be determined. The strategy was successfully applied in the determination of a membrane protein structure, the CopA Cu$^+$-ATPase, when other methods had failed to resolve the heavy atom substructure. MRPM is particularly suited for proteins undergoing large conformational changes where multiple search models should be generated, and it enables the identification of weak but correct molecular replacement solutions with maximum contrast to prime experimental phasing efforts.

## 1. INTRODUCTION

To solve a structure from a macromolecular crystal the phase-problem must be solved. For isomorphous replacement and anomalous scattering methods (in this paper called experimental phasing in unison), phasing can be considered a two-step procedure where initially the heavy atom (HA) substructure is solved, after which the substructure is used to calculate phases for the entire macromolecular structure (Hendrickson, 1991; Dauter et al., 2002). If the substructure is solved, reasonable experimental maps can often be generated from surprisingly weak data thanks to improvements in phase calculation and density modification procedures (e.g. Terwilliger, 2000; Terwilliger, 2001; McCoy, 2002; Jenni et al., 2006; Maier et al., 2006; Keller et al., 2006; Liu, et al., 2011; Li and Li, 2011; Abrescia et al., 2011)

Typically the heavy atom substructure is found using Patterson-based or (less frequently) direct methods (Hendrickson and Ogata, 1997; Weeks and Miller, 1999; Grosse-Kunstleve and Adams, 2003; Sheldrick, 2008; Burla et al., 2003). Such heavy-atom site identification is non-trivial when only weak diffraction data of poor quality are available and often complicated by crystal and data pathologies such as radiation damage and severe anisotropy.

Molecular replacement (MR) is an alternative method for obtaining phase estimates. However, if the experimental data is low resolution and low quality, the end-result will be highly biased by the model (Read, 1986; DeLaBarre and Brunger, 2006), hiding novel features in the structure and preventing rebuilding and refinement of the target structure.

Nonetheless, MR is still useful in such difficult cases. By using molecular replacement at low resolution, an initial starting model, despite very low sequence identity, can generate phases which allow for the identification of HA peaks in anomalous difference Fourier maps. After positioning of the heavy atom(s), the model-biased MR phases can be discarded and phase calculation and improvement conducted using traditional methods. This approach has been used in a number of cases to solve difficult structures (Pedersen et al., 2010).

Here we present a systematic expansion of this methodology that we developed during our work to solve the structure of the CopA Cu$^+$-ATPase (Gourdon et al., 2011a). Identification of heavy atom sites in CopA HA-derivative data turned out to be highly challenging. While an extensive effort was put into the generation of improved derivative and native crystals, a strategy to systematically screen MR-parameters was developed, dubbed Molecular Replacement Parameter Matrix (MRPM) search, since more traditional methods consistently failed.

## 2. MATERIALS AND METHODS

### 2.1 Sample description

CopA is a membrane protein from *Legionella pneumophila* which belong to the well-studied family of primary transporters known as P-type ATPases (Møller et al., 1996; Axelsen and Palmgren, 1998; Møller et al., 2010).

This family has a common core of six transmembrane (TM) helices called the M16 domain, and three soluble domains, known as the A, N and P domains (Morth et al. 2011). Crystallization of CopA resulted in crystals that diffracted to 3.2 Å in the best case, suffering from severe non-isomorphism between most datasets (Supplementary Table 1) (Gourdon et al., 2011a; Gourdon et al., 2011b).

### 2.2 Method description

The identification of a correct MR solution is not trivial when the search model and/or experimental data are of poor quality. Searching using various high resolution data cutoffs and various estimated root mean square coordinate error (r.m.s.) of the search model and using search-models which encompass as much of the asymmetric unit as possible can help (Pedersen et al., 2010).

If the conformational flexibility of the target is cause for concern, a number of different conformational states should also be tested.

Here we test a number of model-conformations and search parameters in a systematic fashion to maximize the searched MR-parameter space. Since the end-goal is to identify unambiguous HA peaks, the numerous MR solutions are scored using this criterion and the corresponding Z-score simultaneously, to help separate correct solutions from noise.

### 2.3 Hardware and software used

The computer used was a regular linux desktop computer (4x Intel Xeon CPU W3540 (2.93GHz), 24G RAM). A total of 397 CPU hours were used for this analysis. In real time the calculations took 4d 3h 20m.

All scripts were made using the Bourne shell (sh). Example scripts sufficient to perform a similar analysis are provided as supplementary material. Programs used were *Phaser*, *PEAKMAX*, *SCALEIT*, *FFT*, *SUPERPOSE*, *pymol* and *gnuplot* (Howell and Smith, 1992; McCoy et al., 2007; McCoy et al., 1994; Ten Eyck, 1973; Krissinel and Henrick, 2004; DeLano; Williams and Kelley, 1993).

## 3. RESULTS AND DISCUSSION

A schematic representation of the MRPM strategy is shown in Figure 1. Manually analyzing the heavy-atom derivative datasets collected, one $K_2PtCl_6$ derivate dataset was identified as the 'best' HA-dataset, i.e. having the most significant anomalous difference signal extending to 5.5 Å resolution (Supplementary Table 2). The strategy was designed to evaluate if MR phases could identify significant anomalous difference peaks in this Pt-derivate dataset.

### 3.1 Generation of the search model library

To obtain a useful library of search models we regard a search model to be composed of a number of *domains* arranged according to an overall *scaffold* representing different conformational states. To further increase the set of covered models the domains may be subjugated to *truncations* of loop regions and *pruning* of the side chain atoms, leading to a library of related search models.

Several full length P-type ATPase structures (mainly of the $Ca^{2+}$-ATPase) are available in the Protein Data Bank (PDB) representing a spectrum of conformational states. For scaffolds, 33 P-type ATPase structures were downloaded and a RMS deviation matrix of the C(alpha) atoms was calculated (Supplementary Table 3). Redundant scaffolds were thus identified, resulting in 15 scaffolds with more than 1 Å r.m.s. deviation from each other (Supplementary Table 4). Models of the soluble A, N and P domains were identified by BLAST, as homologous structures with high sequence identity exist. For the M16 domain, the equivalent part of each of the 15 scaffolds was used. These 4 domains together cover ~71% of the CopA sequence (Supplementary Table 5). Missing parts of CopA included the heavy metal binding domain and two initial TM helices; both are specific features of heavy metal pumps and had unknown positions relative to the scaffolds.

The 4 domains were placed by superposition into the 15 scaffolds, resulting in 15 starting models representing the conformational variability observed in P-type ATPases structures (Supplementary Figure 1 step 1 and 2, Supplementary Figure 2).

To increase signal-to-noise ratio in the MR search it is beneficial to leave out unordered and flexible regions and otherwise incorrect sections of the search model. We tested one full length and three truncated versions (A, N and M16 domain removed respectively) for each starting model (Supplementary Figure 1 step 3). These four versions of each starting model were created in two forms; either all atoms present or pruned to poly-alanines only (Supplementary Figure 1 step 4).

The final search model library contained 120 different search models (Supplementary Table 6-9).

### 3.2 Setting up the MR parameter-matrix search

Six native datasets were selected, based on merits such as good quality of the low resolution data, highest obtained resolution, and best scaling overall to the Pt-derivate dataset (Supplementary Table 2). The solvent-content was calculated to be 62% suggesting one monomer per asymmetric unit.

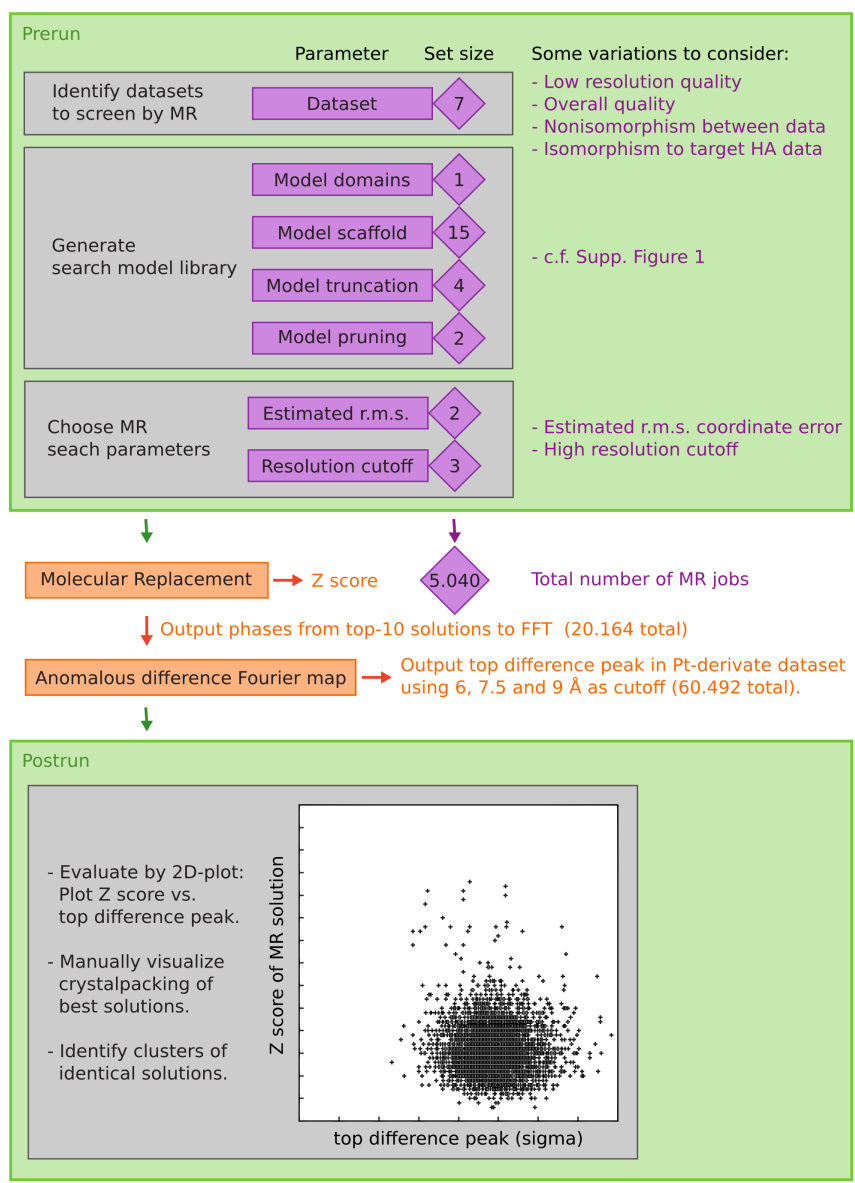

**FIGURE 1. Overview of the MRPM search strategy.** Prerun considerations (top green box) have to be made to identify parameters (dimensions) and sets of values to test for each parameter. The parameters and set size for each parameter shown here are specific for the CopA case. After each MR and FFT calculation, the result is plotted in a 2D-plot to identify clusters of MR solutions having both high Z-score and generating big difference peaks in the Pt-derivate dataset.

Based on previous experience with MR in low quality data (Pedersen et al., 2010), we tested different values for the expected r.m.s. coordinate error (2 or 3 Å) and high resolution limit of the data (4, 6, 8 Å), and left other parameters constant.

The final parameter-matrix contained six parameters-setups for seven datasets using 120 search models, resulting in 5040 MR-searches (Figure 1). As a correct solution was expected to be weak, the ten best final solutions from each run were saved and evaluated. Post-run analysis show that a total of 20.164 suggested MR-solutions were output from the 5.040 MR-searches.

### 3.3 Evaluation

An anomalous difference Fourier map of the derivate Pt-dataset was calculated for each of the 20.164 MR-solutions. Weak peaks in such maps are very sensitive to resolution cutoff so three different cut-offs (6, 7.5, 9Å) were used. The highest difference peak for each of the 60.492 maps was identified and plotted as a function of the Z-score of the input MR solution.

The majority of MR solutions have low Z-scores (<5.5) and do not give rise to significant difference peaks (<5 σ) indicating failed MR searches. However a number of attractive MR-solutions are apparent and through evaluation according to the various screened parameters a tantalizing pattern emerges (Figure 2).

A broad selection of top-scoring solutions was manually analyzed and we found that 30 of these were virtually identical and all identifying the same difference peak (highlighted in Figure 2). All of these required the exclusion of high resolution data, an E2-type scaffold and a poly-alanine model excluding the TM16 domain, and depending on the dataset, the parameters would either give a conspicuous Z-score *or* a conspicuous difference peak.

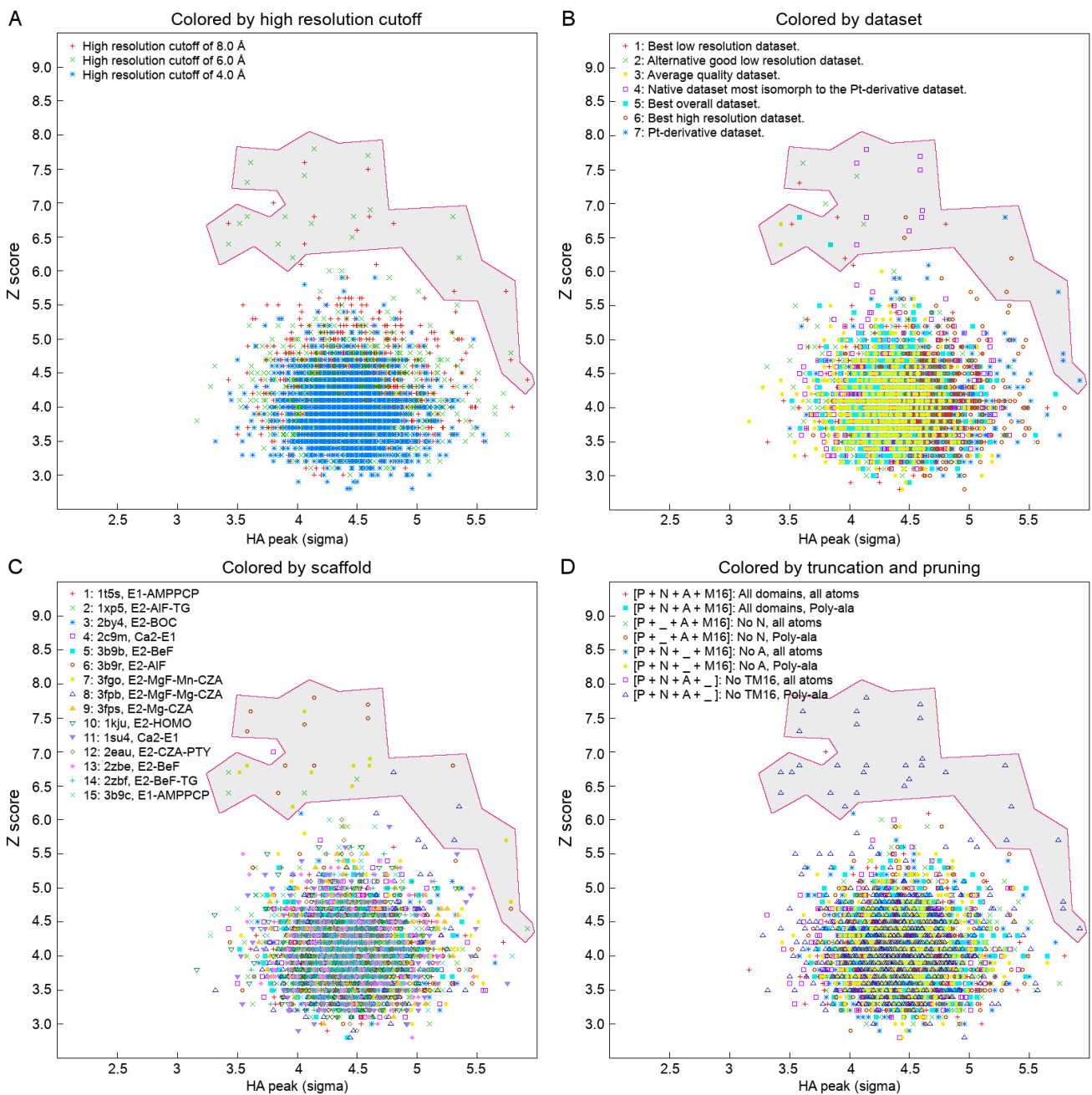

**FIGURE 2. 2D-plot of the result of the MR parameter search.** All solutions are plotted as a function of Z-score and corresponding highest difference peak in the Pt-derivate dataset. The grey area highlights the MR-solutions that turned out to be identical and correct. **A:** High resolution cutoff. **B:** Dataset used. **C:** Scaffold used. The pbd-id is noted, as well as the catalytic state of the structure. **D:** Truncation and pruning used.

The phases from MR using these parameters allowed computation of two initial positions of Pt-atoms leading to experimental phases and the structure to be finally solved.

The best MR-solution as evaluated be Z-score alone (z-score: 7.8) was a correct solution, but the Pt-peak calculated using phases from this particular solution was insignificant (4.14 σ) likely due to non-isomorphism to the Pt-derivative dataset. We must emphasize that even if by serendipity the best possible selection of parameters tested here had been used in a single MR-run, the result would still not be sufficiently clear to be sure of its correctness. Only by comparing a number of solutions does a consistent picture emerge which lends confidence to the further analysis. As two examples of this, one particular solution had a Z-score of 7.0 and another produced a difference peak at 5.79 σ and both turned out to be wrong (Figure 2).

## 4. CONCLUDING REMARKS

For CopA, the presented Molecular Replacement Parameter Matrix search was the only way to initiate phasing. We believe the MRPM search strategy is of general interest for numerous projects with analogous challenges as well as in several more standard molecular replacement applications. It can easily be extended to use more dimensions than presented here. Employing an array of different domains (e.g. domains solved from different organisms) is one example. Testing more datasets, using alternative ways of pruning or even full mutagenesis to the target sequence, are other obvious choices. Furthermore, multiple derivate datasets could be employed to identify different HA-peaks.

Keeping in mind the advent of improved protein-folding algorithms (Rigden et al., 2008; Qian et al., 2007; DiMaio et al., 2011; Bunkóczi and Read, 2011), generic search models (Strop 2007) and automated procedures (Stokes-Rees and Sliz, 2010; Keegan and Winn, 2007), the vital importance of testing different conformational states is accentuated by the work presented here, and it emphasizes an aspect of modeling not currently addressed by *in silico* modeling.

In general, systematic MR-searches are preferential to single MR runs - using the MRPM strategy described here in conjunction with powerful approaches such as MrBUMP and Wide Search Molecular Replacement for instance (Stokes-Rees and Sliz, 2010; Keegan and Winn, 2007). Even if derivative data is not available, a systematic search is more likely to help identify a correct solution and distinguish it from false positives when only low-quality data is available.

MRPM search is CPU-cheap and it is straightforward to implement as scripts (as shown here). We strongly advocate the incorporation of such a strategy directly into the code of MR programs (e.g. Phaser) and/or MR 'black box'-wrappers like MrBUMP.

## Acknowledgements

We are grateful to Gregers R. Andersen for discussions about molecular replacement strategies. We thank Robert M. Stroud and Janet Finer-Moore for reading and reviewing the final manuscript. B.P.P. is supported by a postdoc fellowship from the Carlsberg Foundation and a fellowship from the Danish Cancer Society, P.G. is supported by the Swedish Research Council, X-Y.L. by the China Scholarship Council and P.N. is supported by an advanced research grant (BIOMEMOS) of the European Research Council.

## Author contributions

All authors contributed equally.

## References

Abrescia, N. G. A., Grimes, J. M., Oksanen, H. M., Bamford, J. K. H., Bamford, D. H., et al. (2011). The use of low-resolution phasing followed by phase extension from 7.6 to 2.5 Å resolution with noncrystallographic symmetry to solve the structure of a bacteriophage capsid protein. *Acta Crystallographica. Section D, Biological Crystallography*, *67*, 228-232.

Axelsen, K. B., & Palmgren, M. G. (1998). Evolution of substrate specificities in the P-type ATPase superfamily. *J Mol Evol*, *46*, 84-101.

Bunkóczi, G., & Read, R. J. (2011). Improvement of molecular-replacement models with Sculptor. *Acta Crystallographica. Section D, Biological Crystallography*, *67*, 303-312.

Burla, M. C., Carrozzini, B., Cascarano, G. L., Giacovazzo, C., & Polidori, G. (2003). SAD or MAD phasing: location of the anomalous scatterers. *Acta Crystallographica. Section D, Biological Crystallography*, *59*, 662-669.

Dauter, Z., Dauter, M., & Dodson, E. (2002). Jolly SAD. *Acta Crystallographica. Section D, Biological Crystallography*, *58*, 494-506.

DeLaBarre, B., & Brunger, A. T. (2006). Considerations for the refinement of low-resolution crystal structures. *Acta Crystallogr D Biol Crystallogr*, *62*, 923-32.

DeLano, W. L. The PyMOL Molecular Graphics System. *http://www.pymol.org/*.

DiMaio, F., Terwilliger, T. C., Read, R. J., Wlodawer, A., Oberdorfer, G., et al. (2011). Improved molecular replacement by density- and energy-guided protein structure optimization. *Nature*, *473*, 540-543.

Gourdon, P., Liu, X.-Y., Skjørringe, T., Morth, J. P., Møller, L. B., et al. (2011a). Crystal structure of a copper-transporting PIB-type ATPase. *Nature*, *475*, 59-64.

Gourdon, P., Andersen, J. L., Hein, K. L., Bublitz, M., Pedersen, B. P., et al. (2011b). HiLiDe—Systematic Approach to Membrane Protein Crystallization in Lipid and Detergent. *Crystal Growth & Design*, *11*, 2098-2106.

Grosse-Kunstleve, R. W., & Adams, P. D. (2003). Substructure search procedures for macromolecular structures. *Acta Crystallographica. Section D, Biological Crystallography*, *59*, 1966-1973.

Hendrickson, W. A. (1991). Determination of macromolecular structures from anomalous diffraction of synchrotron radiation. *Science (New York, N.Y.)*, *254*, 51-58.

Hendrickson, W. A., & Ogata, C. M. (1997). Phase determination from multiwavelength anomalous diffraction measurements. In *Macromolecular Crystallography Part A* (pp. 494-523). Academic Press.

Howell, P. L., & Smith, G. D. (1992). Identification of heavy-atom derivatives by normal probability methods. *Journal of Applied Crystallography*, *25*, 81-86.

Jenni, S., Leibundgut, M., Maier, T., & Ban, N. (2006). Architecture of a fungal fatty acid synthase at 5 A resolution. *Science (New York, N.Y.)*, *311*, 1263-1267.

Keegan, R. M., & Winn, M. D. (2007). Automated search-model discovery and preparation for structure solution by molecular replacement. *Acta Crystallographica Section D Biological Crystallography*, *63*, 447-457.

Keller, S., Pojer, F., Heide, L., & Lawson, D. M. (2006). Molecular replacement in the "twilight zone": structure determination of the non-haem iron oxygenase NovR from Streptomyces spheroides through repeated density modification of a poor molecular-replacement solution. *Acta Crystallographica. Section D, Biological Crystallography*, *62*, 1564-1570.

Krissinel, E., & Henrick, K. (2004). Secondary-structure matching (SSM), a new tool for fast protein structure alignment in three dimensions. *Acta Crystallographica Section D Biological Crystallography*, *60*, 2256-2268.

Li, W., & Li, F. (2011). Cross-crystal averaging with search models to improve molecular replacement phases. *Structure (London, England: 1993)*, *19*, 155-161.

Liu, Q., Zhang, Z., & Hendrickson, W. A. (2011). Multi-crystal anomalous diffraction for low-resolution macromolecular phasing. *Acta Crystallographica. Section D, Biological Crystallography*, *67*, 45-59.

Maier, T., Jenni, S., & Ban, N. (2006). Architecture of mammalian fatty acid synthase at 4.5 A resolution. *Science (New York, N.Y.)*, *311*, 1258-1262.

McCoy, A. J. (2002). New applications of maximum likelihood and Bayesian statistics in macromolecular crystallography. *Current Opinion in Structural Biology*, *12*, 670-673.

McCoy, A. J., Grosse-Kunstleve, R. W., Adams, P. D., Winn, M. D., Storoni, L. C., et al. (2007). *Phaser* crystallographic software. *Journal of Applied Crystallography*, *40*, 658-674.

Møller, J. V., Olesen, C., Winther, A.-M. L., & Nissen, P. (2010). The sarcoplasmic Ca2+-ATPase: design of a perfect chemi-osmotic pump. *Quarterly Reviews of Biophysics*, 1-66.

Møller, J. V., Juul, B., & le Maire, M. (1996). Structural organization, ion transport, and energy transduction of P-type ATPases. *Biochim Biophys Acta*, *1286*, 1-51.

Morth, J. P., Pedersen, B. P., Toustrup-Jensen, M. S., Sorensen, T. L., Petersen, J., et al. (2007). Crystal structure of the sodium-potassium pump. *Nature*, *450*, 1043-9.

Morth, J. P., Pedersen, B. P., Buch-Pedersen, M. J., Andersen, J. P., Vilsen, B., et al. (2011). A structural overview of the plasma membrane Na+,K+-ATPase and H+-ATPase ion pumps. *Nature Reviews. Molecular Cell Biology*, *12*, 60-70.

Pedersen, B. P., Buch-Pedersen, M. J., Morth, J. P., Palmgren, M. G., & Nissen, P. (2007). Crystal structure of the plasma membrane proton pump. *Nature*, *450*, 1111-1114.

Pedersen, B. P., Morth, J. P., & Nissen, P. (2010). Structure determination using poorly diffracting membrane protein crystals - Lessons from the H+ and Na+,K+-ATPases. *Acta Crystallographica Section D Biological Crystallography*, *66*, 309-313.

Qian, B., Raman, S., Das, R., Bradley, P., McCoy, A. J., et al. (2007). High-resolution structure prediction and the crystallographic phase problem. *Nature*, *450*, 259-64.

Read, R. J. (1986). Improved Fourier coefficients for maps using phases from partial structures with errors. *Acta Crystallographica Section A Foundations of Crystallography*, *42*, 140-149.

Rigden, D. J., Keegan, R. M., & Winn, M. D. (2008). Molecular replacement using ab initio polyalanine models generated with ROSETTA. *Acta Crystallographica. Section D, Biological Crystallography*, *64*, 1288-1291.

Sheldrick, G. M. (2008). A short history of SHELX. *Acta Crystallogr A*, *64*, 112-22.

Stokes-Rees, I., & Sliz, P. (2010). Protein structure determination by exhaustive search of Protein Data Bank derived databases. *Proceedings of the National Academy of Sciences*, *107*, 21476-21481.

Strop, P., Brzustowicz, M. R., & Brunger, A. T. (2007). Ab initio molecular-replacement phasing for symmetric helical membrane proteins. *Acta Crystallographica. Section D, Biological Crystallography*, *63*, 188-196.

Ten Eyck, L. F. (1973). Crystallographic fast Fourier transforms. *Acta Crystallographica Section A*, *29*, 183-191.

Terwilliger, T. C. (2001). Map-likelihood phasing. *Acta Crystallographica. Section D, Biological Crystallography*, *57*, 1763-1775.

Terwilliger, T. C. (2000). Maximum-likelihood density modification. *Acta Crystallographica. Section D, Biological Crystallography*, *56*, 965-972.

The CCP4 suite: programs for protein crystallography. (1994).*Acta Crystallographica. Section D, Biological Crystallography*, *50*, 760-763.

Weeks, C. M., & Miller, R. (1999). Optimizing Shake-and-Bake for proteins. *Acta Crystallographica. Section D, Biological Crystallography*, *55*, 492-500.

Williams, T., & Kelley, C. (1993). GNUPLOT: An Interactive Plotting Program. *www.gnuplot.info*

**SUPPLEMENTARY MATERIAL**

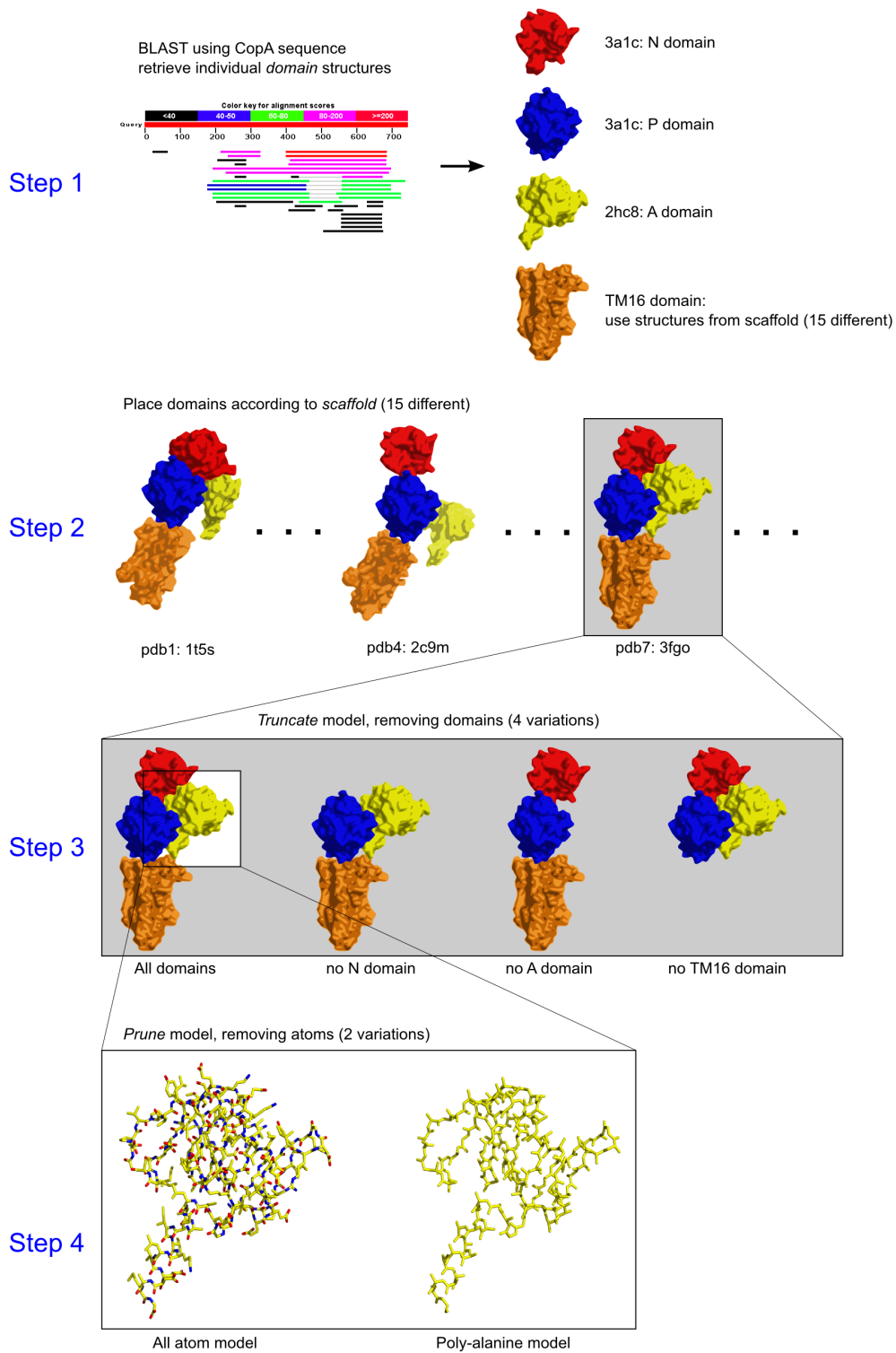

Final library contains 120 search models.

**Supplementary Figure 1. Generation of the search model library.** Generation of the library is divided into 4 steps. *Step 1:* Identify the domains to use. *Step 2:* Identify the scaffolds used to place the domains into representing different possible conformations and place the domains into these scaffolds. *Step 3:* Identify and generate a number of truncations removing different domains, since one incorrectly placed domain can make the difference between success and failure. *Step 4:* Prune the atoms of the models to generate variations. In this particular case only two different pruning schemes were used: all atom or reduction to poly-alanine.

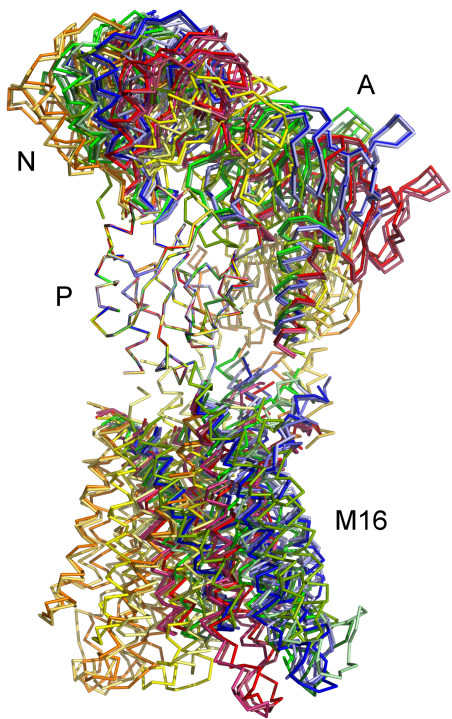

**Supplementary Figure 2. Superposition of the 15 starting models after step 2 in Supp. Figure 1.** All models are superposed on the P domain and each model has a distinct color. The conformational variation obtained by using different scaffolds is evident. The functional cycles of P-type ATPases is characterized by four principal conformations (Møller et al., 2010). The colors are chosen to emphasize that the models fall into these classes: Red shades are the Occluded E2-Pi forms. Blue shades are the occluded E2 transition state forms. Green shades are the open E2 forms. Yellow shades are the E1 forms.

**Supplementary Table 1. Non-isomorphism between datasets.** Total R-factor on F by *SCALEIT* from 50-6 Å. Dataset 7 is the Pt-derivative dataset (cf. Supplementary Table 2).

| R(cross) | Dataset 1 | Dataset 2 | Dataset 3 | Dataset 4 | Dataset 5 | Dataset 6 | Dataset 7 |
|---|---|---|---|---|---|---|---|
| **Dataset 1** | - | 0.078 | 0.342 | 0.192 | 0.094 | 0.213 | 0.355 |
| **Dataset 2** | 0.078 | - | 0.367 | 0.152 | 0.138 | 0.259 | 0.335 |
| **Dataset 3** | 0.342 | 0.367 | - | 0.407 | 0.329 | 0.300 | 0.499 |
| **Dataset 4** | 0.192 | 0.152 | 0.407 | - | 0.243 | 0.338 | 0.297 |
| **Dataset 5** | 0.094 | 0.138 | 0.329 | 0.243 | - | 0.171 | 0.436 |
| **Dataset 6** | 0.213 | 0.259 | 0.300 | 0.338 | 0.171 | - | 0.439 |
| **Dataset 7** | 0.355 | 0.335 | 0.499 | 0.297 | 0.436 | 0.439 | - |

**Supplementary Table 2. Dataset statistics.**

| | Dataset 1 | Dataset 2 | Dataset 3 | Dataset 4 | Dataset 5 | Dataset 6 | Dataset 7 |
|---|---|---|---|---|---|---|---|
| Type | native | native | native | native | native | native | K2PtCl-derivative |
| Reason for inclusion in this analysis. | High quality at low resolution. | Alternative high quality at low resolution. | Average quality. | Native most isomorph to the Pt-dataset. | High quality overall. | Best high resolution. | Best heavy atom derivative. |
| **Overall statistics** | | | | | | | |
| Space group | $P2_12_12_1$ | $P2_12_12_1$ | $P2_12_12_1$ | $P2_12_12_1$ | $P2_12_12_1$ | $P2_12_12_1$ | $P2_12_12_1$ |
| Cell dimensions | | | | | | | |
| a, b, c (Å) | 44.21, 72.51, 328.92 | 44.12, 72.59, 330.18 | 43.77, 71.76, 324.05 | 44.72, 72.61, 329.53 | 44.53, 72.76, 329.65 | 43.87, 72.14, 327.28 | 45.03, 72.53, 329.25 |
| $\alpha$, $\beta$, $\gamma$ (°) | 90, 90, 90 | 90, 90, 90 | 90, 90, 90 | 90, 90, 90 | 90, 90, 90 | 90, 90, 90 | 90, 90, 90 |
| Resolution | 50-3.7 (3.8-3.7)[a] | 50-4.0 (4.1-4.0) | 50-3.7 (3.8-3.7) | 50-4.1 (4.2-4.1) | 50-3.55 (3.6-3.55) | 50-3.4 (3.5-3.4) | 50-5.5 (6.0-5.5) |
| Rsym (%) | 15.4 (92.8) | 13.7 (68.4) | 32.2 (154.2) | 12.1 (36.9) | 14.4 (161.7) | 10.7 (117.3) | 10.9 (54.9) |
| Rmeas (%)[b] | 16.6 (100.0) | 15.1 (75.1) | 34.8 (169.2) | 15.0 (46.1) | 15.4 (171.5) | 11.9 (132.7) | 12.5 (62.7) |
| Rmrgd-F (%)[c] | 18.3 (74.5) | 19.4 (72.3) | 37.9 (156.6) | 25.9 (69.6) | 17.0 (109.9) | 16.7 (116.3) | 18.4 (63.3) |
| I/sig(I) | 12.65 (2.05) | 11.57 (2.41) | 6.2 (1.00) | 6.63 (2.52) | 11.77 (1.46) | 12.71 (1.55) | 9.28 (2.65) |
| Completeness (%) | 99.7 (99.9) | 99.4 (99.5) | 99.2 (97.2) | 99.1 (99.8) | 99.8 (99.3) | 98.7 (94.9) | 99.0 (98.4) |
| Redundancy | 7.0 (7.2) | 5.6 (5.7) | 6.8 (5.1) | 2.8 (2.7) | 8.6 (8.8) | 5.4 (4.3) | 4.3 (4.2) |
| SigAno[d] | - | - | - | - | - | - | 1.181 (0.805) |
| **High resolution quality** | | | | | | | |
| Rmrgd-F (4.1-4.2 Å) (%) | 30.7 | 48.4 | 74.4 | 69.6 | 23.4 | 17.7 | - |
| I/sig(I) (4.1-4.2 Å) | 5.28 | 3.33 | 2.13 | 2.52 | 6.66 | 8.94 | - |
| **Low resolution quality** | | | | | | | |
| Rmrgd-F (20-30 Å) (%) | 1.2 | 1.1 | 2.1 | 3.2 | 2.0 | 2.0 | 3.8 |
| I/sig(I) (20-30 Å) | 66.69 | 56.45 | 33.93 | 20.33 | 38.80 | 48.02 | 20.68 |
| **Pt-derivate signal** | | | | | | | |
| SigAno (15-50 Å) | - | - | - | - | - | - | 2.980 |
| SigAno (7-8 Å) | - | - | - | - | - | - | 1.158 |

a: Values in parenthesis are from the highest resolution shell.
b: R-meas = redundancy independent R-factor (intensities). (Diederichs & Karplus (1997), Nature Struct. Biol. 4, 269-275.)
c: Rmrgd-F = quality of amplitudes (F) in the scaled data set. (Diederichs & Karplus (1997), Nature Struct. Biol. 4, 269-275.)
d: SigAno = mean anomalous difference in units of its estimated standard deviation ($|F(+)-F(-)|/Sigma$). F(+), F(-) are structure factor estimates obtained from the merged intensity observations in each parity class (as calculated in XSCALE).

**Supplementary Table 3. C-alpha r.m.s. deviation between the 32 initially identified Ca²⁺-ATPase scaffolds.** The code refers to the scaffolds pdb-id. Red notes r.m.s.d. below 1 Å.

**Supplementary Table 4. Pruned C-alpha r.m.s. deviation between the initially identified Ca2+-ATPase scaffolds.** Models with a deviation less than 1 Å were removed from Supplementary table 3. The resulting list shown here was used for the initial 14 scaffolds to generate the search library. The code refers to the scaffolds pdb-id.

|        | 1T5S | 1XP5 | 2BY4 | 2C9M | 3B9B | 3B9R | 3FGO | 3FPB | 3FPS | 1KJU | 1SU4 | 2EAU | 2ZBE | 2ZBF |
|--------|------|------|------|------|------|------|------|------|------|------|------|------|------|------|
| **1T5S** | 0.00 | 4.56 | 3.63 | 2.14 | 3.90 | 4.74 | 4.53 | 4.88 | 3.85 | 4.25 | 2.29 | 3.84 | 4.27 | 4.31 |
| **1XP5** | 4.55 | 0.00 | 2.65 | 3.44 | 3.48 | 1.54 | 1.50 | 1.34 | 2.84 | 2.79 | 2.39 | 2.83 | 2.83 | 2.64 |
| **2BY4** | 3.63 | 2.67 | 0.00 | 4.36 | 3.83 | 2.83 | 2.92 | 3.00 | 1.03 | 3.56 | 4.51 | 1.38 | 3.48 | 2.85 |
| **2C9M** | 2.14 | 3.43 | 4.47 | 0.00 | 3.11 | 3.57 | 2.84 | 2.95 | 4.19 | 4.18 | 2.57 | 4.24 | 2.10 | 3.09 |
| **3B9B** | 3.89 | 3.50 | 3.73 | 3.11 | 0.00 | 3.09 | 3.11 | 3.17 | 3.50 | 4.12 | 2.53 | 3.35 | 1.79 | 1.82 |
| **3B9R** | 4.65 | 1.54 | 2.79 | 3.57 | 3.11 | 0.00 | 1.08 | 1.09 | 2.87 | 3.20 | 2.41 | 3.26 | 2.59 | 2.26 |
| **3FGO** | 4.50 | 1.50 | 2.99 | 2.84 | 3.15 | 1.09 | 0.00 | 1.00 | 2.94 | 2.97 | 2.42 | 3.24 | 2.63 | 2.19 |
| **3FPB** | 4.89 | 1.34 | 3.00 | 2.95 | 3.17 | 1.09 | 1.00 | 0.00 | 2.92 | 2.86 | 2.45 | 3.21 | 2.62 | 2.37 |
| **3FPS** | 3.84 | 2.86 | 1.03 | 4.19 | 3.48 | 2.87 | 2.94 | 2.93 | 0.00 | 3.58 | 4.17 | 1.10 | 3.00 | 2.57 |
| **1KJU** | 4.23 | 2.80 | 3.56 | 4.14 | 4.08 | 3.19 | 3.02 | 2.86 | 3.50 | 0.00 | 2.97 | 3.50 | 3.75 | 3.68 |
| **1SU4** | 2.30 | 2.39 | 4.51 | 2.57 | 2.53 | 2.42 | 2.47 | 2.42 | 4.17 | 2.97 | 0.00 | 4.48 | 3.69 | 1.92 |
| **2EAU** | 3.86 | 2.83 | 1.39 | 4.24 | 3.35 | 3.23 | 3.24 | 3.21 | 1.11 | 3.50 | 4.48 | 0.00 | 3.39 | 3.18 |
| **2ZBE** | 4.18 | 2.83 | 3.51 | 2.05 | 1.79 | 2.59 | 2.64 | 2.62 | 2.98 | 3.74 | 3.69 | 3.41 | 0.00 | 1.58 |
| **2ZBF** | 4.25 | 2.60 | 2.83 | 3.09 | 1.83 | 2.24 | 2.21 | 2.36 | 2.58 | 3.68 | 1.92 | 3.16 | 1.58 | 0.00 |

**Supplementary Table 5. Models identified by BLAST, used to generate the domains for the search model library.**

| Domain | pdb | CopA coverage (%) | seq. id to CopA (%) | C(alpha) r.m.s.d. to final CopA model |
|--------|-----|-------------------|---------------------|----------------------------------------|
| A | 2hc8 | 102 / 732 (13.9%) | 57 / 102 (55.9%) | 1.06 Å |
| N | 3a1c | 119 / 732 (16.3%) | 44 / 119 (37.0%) | 1.55 Å |
| P | 3a1c | 146 / 732 (19.9%) | 75 / 146 (51.4%) | 1.48 Å |
| M16 (Ca²⁺-ATPase) | 14 different[a] | 152 / 732 (20.7%) | 41 / 152 (27.0%) | 2.58 Å - 4.12 Å |
| M16 (H⁺-ATPase) | 3b8c | 185 / 732 (25.3%) | 38 / 185 (20.5%) | 3.06 Å |
| Ca²⁺-ATPase (P+N+A+M16) | - | 519 / 732 (70.9%) | 217 / 519 (41.8%) | - |
| H⁺-ATPase (P+N+A+M16) | - | 552 / 732 (75.4%) | 214 / 552 (38.8%) | - |

a) For the M16 domain, the relevant part of the individual scaffolds were used, since M16 displays considerable intra-domain variation (cf. Supplementary Table 4 for a list of these scaffolds).

**Supplementary Table 6. r.m.s.d. of search-models with no truncation, i.e. all domains (A+N+P+M16).** Pdbs are numbered 1-15. Their order is identical to the order seen in Figure 2C and Supp. Table 4, with the addition of the H+-ATPase (3b8c) added as a scaffold (pdb15). Red notes r.m.s.d. below 1 Å.

| | pdb1 | pdb2 | pdb3 | pdb4 | pdb5 | pdb6 | pdb7 | pdb8 | pdb9 | pdb10 | pdb11 | pdb12 | pdb13 | pdb14 | pdb15 |
|---|---|---|---|---|---|---|---|---|---|---|---|---|---|---|---|
| **pdb1** | 0.00 | 4.24 | 3.67 | 2.68 | 3.93 | 4.09 | 3.98 | 4.23 | 3.80 | 4.43 | 2.29 | 3.69 | 3.67 | 3.92 | 3.04 |
| **pdb2** | 4.24 | 0.00 | 2.82 | 2.11 | 3.15 | 1.37 | 1.41 | 1.91 | 2.94 | 2.84 | 2.09 | 3.05 | 2.70 | 2.47 | 3.91 |
| **pdb3** | 3.67 | 2.82 | 0.00 | 5.15 | 3.97 | 2.88 | 3.14 | 3.02 | 0.94 | 3.26 | 4.12 | 1.13 | 3.77 | 3.04 | 3.66 |
| **pdb4** | 2.68 | 2.11 | 5.15 | 0.00 | 1.97 | 2.15 | 2.88 | 2.93 | 3.72 | 3.02 | 2.82 | 5.16 | 2.84 | 2.14 | 3.43 |
| **pdb5** | 3.92 | 3.15 | 3.98 | 1.97 | 0.00 | 2.75 | 2.91 | 3.34 | 3.97 | 4.07 | 2.11 | 4.10 | 1.39 | 1.73 | 4.40 |
| **pdb6** | 4.09 | 1.37 | 2.87 | 2.15 | 2.74 | 0.00 | 0.92 | 1.57 | 3.12 | 3.23 | 2.55 | 3.20 | 2.39 | 1.87 | 4.74 |
| **pdb7** | 4.17 | 1.41 | 3.14 | 2.88 | 2.94 | 0.92 | 0.00 | 1.41 | 3.27 | 3.21 | 2.35 | 3.35 | 2.53 | 2.07 | 3.74 |
| **pdb8** | 4.23 | 1.91 | 3.26 | 2.93 | 3.31 | 1.63 | 1.41 | 0.00 | 3.23 | 3.52 | 3.20 | 3.35 | 2.98 | 2.62 | 3.85 |
| **pdb9** | 3.83 | 2.94 | 0.94 | 3.72 | 3.97 | 3.12 | 3.08 | 3.23 | 0.00 | 3.27 | 4.03 | 0.94 | 3.71 | 3.29 | 3.64 |
| **pdb10** | 4.43 | 2.83 | 3.26 | 3.02 | 4.00 | 3.23 | 3.21 | 3.52 | 3.27 | 0.00 | 2.87 | 3.25 | 3.63 | 3.31 | 4.89 |
| **pdb11** | 2.29 | 2.09 | 4.12 | 2.82 | 2.11 | 2.55 | 2.35 | 3.18 | 4.03 | 2.87 | 0.00 | 6.44 | 1.76 | 4.66 | 2.53 |
| **pdb12** | 3.82 | 3.05 | 1.13 | 5.16 | 4.10 | 3.05 | 3.39 | 3.39 | 0.94 | 3.25 | 6.42 | 0.00 | 3.99 | 3.38 | 3.54 |
| **pdb13** | 3.67 | 2.70 | 3.77 | 2.84 | 1.39 | 2.39 | 2.53 | 2.98 | 3.98 | 3.62 | 1.76 | 3.99 | 0.00 | 1.71 | 4.30 |
| **pdb14** | 3.83 | 2.47 | 3.06 | 2.13 | 1.73 | 1.87 | 2.08 | 2.62 | 3.29 | 3.31 | 4.52 | 3.31 | 1.65 | 0.00 | 4.58 |
| **pdb15** | 3.02 | 3.88 | 3.68 | 3.56 | 4.41 | 4.74 | 3.86 | 3.86 | 3.64 | 4.74 | 2.51 | 3.55 | 4.04 | 4.09 | 0.00 |

**Supplementary Table 7. r.m.s.d. of search-models with no N domain (A+P+TM16).** Pdbs are numbered 1-15. Their order is identical to the order seen in Figure 2C and Supp. Table 4, with the addition of the H+-ATPase (3b8c) added as a scaffold (pdb15). Red notes r.m.s.d. below 1 Å.

| | pdb1 | pdb2 | pdb3 | pdb4 | pdb5 | pdb6 | pdb7 | pdb8 | pdb9 | pdb10 | pdb11 | pdb12 | pdb13 | pdb14 | pdb15 |
|---|---|---|---|---|---|---|---|---|---|---|---|---|---|---|---|
| **pdb1** | 0.00 | 4.14 | 3.39 | 2.67 | 3.96 | 4.01 | 3.70 | 4.13 | 3.61 | 3.99 | 2.27 | 3.46 | 4.08 | 3.83 | 2.51 |
| **pdb2** | 4.14 | 0.00 | 2.39 | 3.97 | 2.75 | 1.12 | 1.08 | 1.17 | 2.44 | 2.58 | 4.57 | 2.52 | 2.67 | 1.76 | 3.22 |
| **pdb3** | 3.44 | 2.39 | 0.00 | 3.76 | 3.53 | 2.08 | 2.25 | 2.37 | 0.89 | 3.82 | 3.76 | 1.11 | 3.15 | 2.52 | 2.59 |
| **pdb4** | 2.67 | 3.96 | 3.76 | 0.00 | 3.77 | 3.87 | 3.79 | 4.11 | 3.70 | 4.42 | 3.38 | 3.73 | 3.99 | 4.03 | 3.22 |
| **pdb5** | 3.96 | 2.79 | 3.53 | 3.95 | 0.00 | 2.64 | 2.79 | 2.68 | 3.52 | 3.47 | 3.82 | 3.49 | 1.37 | 1.82 | 3.70 |
| **pdb6** | 4.01 | 1.12 | 2.08 | 3.94 | 2.55 | 0.00 | 0.88 | 0.97 | 2.13 | 2.89 | 4.29 | 2.16 | 2.56 | 1.68 | 3.97 |
| **pdb7** | 3.86 | 1.08 | 2.25 | 3.79 | 2.80 | 0.88 | 0.00 | 0.76 | 2.22 | 2.65 | 4.38 | 2.26 | 2.69 | 1.77 | 3.11 |
| **pdb8** | 4.26 | 1.17 | 2.37 | 4.08 | 2.68 | 0.97 | 0.76 | 0.00 | 2.32 | 2.87 | 4.40 | 2.32 | 2.62 | 1.94 | 4.30 |
| **pdb9** | 3.66 | 2.44 | 0.91 | 3.69 | 3.52 | 2.13 | 2.22 | 2.32 | 0.00 | 3.89 | 3.48 | 1.01 | 3.15 | 2.65 | 2.51 |
| **pdb10** | 3.99 | 2.56 | 3.82 | 4.42 | 3.47 | 2.89 | 2.65 | 2.87 | 3.89 | 0.00 | 4.85 | 3.77 | 3.27 | 2.48 | 3.82 |
| **pdb11** | 2.27 | 4.56 | 3.76 | 3.38 | 3.82 | 4.29 | 4.17 | 4.36 | 3.48 | 4.84 | 0.00 | 3.50 | 4.17 | 4.17 | 2.52 |
| **pdb12** | 3.56 | 2.52 | 1.11 | 3.73 | 3.49 | 2.16 | 2.26 | 2.32 | 1.01 | 3.77 | 3.50 | 0.00 | 3.30 | 2.67 | 2.51 |
| **pdb13** | 3.80 | 2.67 | 3.15 | 3.99 | 1.38 | 2.58 | 2.69 | 2.64 | 3.15 | 3.27 | 4.27 | 3.30 | 0.00 | 1.76 | 3.79 |
| **pdb14** | 3.84 | 1.76 | 2.52 | 4.03 | 1.82 | 1.68 | 1.77 | 1.94 | 2.65 | 2.48 | 4.22 | 2.67 | 1.76 | 0.00 | 3.82 |
| **pdb15** | 2.43 | 3.24 | 2.58 | 3.60 | 3.78 | 4.07 | 3.07 | 3.62 | 2.42 | 3.71 | 2.48 | 2.51 | 3.53 | 3.28 | 0.00 |

**Supplementary Table 8. r.m.s.d. of search-models with no A domain (N+P+TM16).** Pdbs are numbered 1-15. Their order is identical to the order seen in Figure 2C and Supp. Table 4, with the addition of the H+-ATPase (3b8c) added as a scaffold (pdb15). Red notes r.m.s.d. below 1 Å.

| | pdb1 | pdb2 | pdb3 | pdb4 | pdb5 | pdb6 | pdb7 | pdb8 | pdb9 | pdb10 | pdb11 | pdb12 | pdb13 | pdb14 | pdb15 |
|---|---|---|---|---|---|---|---|---|---|---|---|---|---|---|---|
| **pdb1** | 0.00 | 4.12 | 3.58 | 1.94 | 3.51 | 3.75 | 3.80 | 4.14 | 3.72 | 4.31 | 1.82 | 3.60 | 3.64 | 3.88 | 2.96 |
| **pdb2** | 4.09 | 0.00 | 2.13 | 2.11 | 3.07 | 1.48 | 1.44 | 2.06 | 2.33 | 2.57 | 2.09 | 2.29 | 2.43 | 2.36 | 3.80 |
| **pdb3** | 3.57 | 2.13 | 0.00 | 5.13 | 3.68 | 2.40 | 2.50 | 2.56 | 1.02 | 2.62 | 5.83 | 1.03 | 3.39 | 2.65 | 3.38 |
| **pdb4** | 1.94 | 2.11 | 5.13 | 0.00 | 1.97 | 2.15 | 2.18 | 2.93 | 4.40 | 3.02 | 1.24 | 5.12 | 2.84 | 2.95 | 2.25 |
| **pdb5** | 3.52 | 3.07 | 3.61 | 1.97 | 0.00 | 2.56 | 2.81 | 3.56 | 3.63 | 3.92 | 2.11 | 3.94 | 1.36 | 1.91 | 4.12 |
| **pdb6** | 3.78 | 1.48 | 2.40 | 2.15 | 2.57 | 0.00 | 0.90 | 1.85 | 2.59 | 3.12 | 2.55 | 2.42 | 2.14 | 1.67 | 3.72 |
| **pdb7** | 3.80 | 1.44 | 2.45 | 2.18 | 2.80 | 0.90 | 0.00 | 1.52 | 2.67 | 3.16 | 2.22 | 2.48 | 2.31 | 1.84 | 3.69 |
| **pdb8** | 4.14 | 2.02 | 2.56 | 2.93 | 3.56 | 1.85 | 1.52 | 0.00 | 2.69 | 3.37 | 3.07 | 2.50 | 2.83 | 2.57 | 3.78 |
| **pdb9** | 3.72 | 2.33 | 1.02 | 4.40 | 3.76 | 2.59 | 2.67 | 2.69 | 0.00 | 2.74 | 5.20 | 0.86 | 3.27 | 2.76 | 3.41 |
| **pdb10** | 4.50 | 2.57 | 2.62 | 3.02 | 3.94 | 3.12 | 3.16 | 3.37 | 2.74 | 0.00 | 2.87 | 2.71 | 3.51 | 3.35 | 4.79 |
| **pdb11** | 1.82 | 2.09 | 5.79 | 1.24 | 2.11 | 2.55 | 2.22 | 3.07 | 5.20 | 2.87 | 0.00 | 6.45 | 2.34 | 4.10 | 2.33 |
| **pdb12** | 3.64 | 2.29 | 1.03 | 5.11 | 3.97 | 2.42 | 2.49 | 2.50 | 0.86 | 2.70 | 6.45 | 0.00 | 3.52 | 2.78 | 3.34 |
| **pdb13** | 3.61 | 2.43 | 3.39 | 2.84 | 1.36 | 2.14 | 2.39 | 2.83 | 3.28 | 3.51 | 2.34 | 3.52 | 0.00 | 1.72 | 4.00 |
| **pdb14** | 3.79 | 2.36 | 2.65 | 2.95 | 1.91 | 1.67 | 1.84 | 2.57 | 2.76 | 3.34 | 4.10 | 2.78 | 1.73 | 0.00 | 3.64 |
| **pdb15** | 2.96 | 3.69 | 3.47 | 2.23 | 4.14 | 3.72 | 3.67 | 3.78 | 3.37 | 4.78 | 2.35 | 3.35 | 4.06 | 3.66 | 0.00 |

**Supplementary Table 9. r.m.s.d. of search-models with no TM16 domain (A+N+P).** Pdbs are numbered 1-15. Their order is identical to the order seen in Figure 2C and Supp. Table 4, with the addition of the H+-ATPase (3b8c) added as a scaffold (pdb15). Red notes r.m.s.d. below 1 Å. Note that without the transmembrane domain present the rmsd drops in may cases (compare Supp. Table 9 to Supp. Table 6-8). Especially pdb2, pdb6 and pdb7 are similar, and pdb3, pdb9 and pdb12 are similar as expected from the conformations that they represent.

| | pdb1 | pdb2 | pdb3 | pdb4 | pdb5 | pdb6 | pdb7 | pdb8 | pdb9 | pdb10 | pdb11 | pdb12 | pdb13 | pdb14 | pdb15 |
|---|---|---|---|---|---|---|---|---|---|---|---|---|---|---|---|
| **pdb1** | 0.00 | 2.01 | 3.40 | 0.97 | 1.66 | 2.02 | 1.96 | 3.01 | 2.96 | 4.07 | 2.86 | 3.16 | 2.46 | 1.73 | 2.97 |
| **pdb2** | 3.54 | 0.00 | 3.17 | 2.09 | 2.01 | 0.58 | 0.56 | 1.66 | 3.50 | 2.66 | 2.01 | 3.31 | 2.07 | 1.98 | 4.18 |
| **pdb3** | 3.40 | 3.17 | 0.00 | 3.90 | 4.19 | 3.38 | 3.52 | 3.66 | 0.59 | 2.98 | 3.88 | 0.53 | 3.85 | 4.25 | 3.62 |
| **pdb4** | 0.90 | 2.09 | 3.90 | 0.00 | 1.21 | 2.10 | 2.11 | 2.95 | 4.21 | 2.90 | 2.82 | 4.06 | 1.37 | 1.61 | 3.38 |
| **pdb5** | 3.53 | 1.99 | 4.00 | 1.21 | 0.00 | 1.90 | 2.02 | 2.90 | 4.54 | 2.57 | 1.24 | 4.32 | 0.75 | 0.59 | 4.10 |
| **pdb6** | 3.12 | 0.58 | 3.38 | 2.10 | 1.90 | 0.00 | 0.54 | 1.68 | 3.51 | 2.68 | 2.00 | 3.37 | 1.94 | 1.71 | 3.56 |
| **pdb7** | 3.09 | 0.56 | 3.52 | 2.11 | 2.02 | 0.54 | 0.00 | 1.40 | 3.73 | 2.66 | 2.15 | 3.59 | 2.06 | 1.82 | 3.36 |
| **pdb8** | 3.01 | 1.65 | 3.66 | 2.93 | 2.90 | 1.68 | 1.40 | 0.00 | 3.83 | 3.07 | 3.03 | 3.53 | 2.91 | 2.72 | 3.91 |
| **pdb9** | 3.03 | 3.50 | 0.59 | 4.21 | 4.51 | 3.51 | 3.73 | 3.83 | 0.00 | 3.23 | 4.23 | 0.45 | 4.03 | 4.57 | 3.39 |
| **pdb10** | 4.56 | 2.65 | 2.98 | 2.90 | 2.53 | 2.68 | 2.66 | 3.07 | 3.11 | 0.00 | 2.79 | 2.92 | 2.64 | 2.54 | 4.57 |
| **pdb11** | 2.86 | 2.01 | 3.88 | 2.82 | 1.24 | 2.00 | 2.15 | 3.03 | 4.23 | 2.79 | 0.00 | 4.06 | 1.63 | 1.73 | 2.02 |
| **pdb12** | 3.17 | 3.31 | 0.53 | 4.06 | 4.32 | 3.37 | 3.59 | 3.20 | 0.45 | 3.20 | 4.06 | 0.00 | 4.02 | 4.28 | 3.44 |
| **pdb13** | 2.46 | 2.07 | 3.85 | 1.37 | 0.75 | 1.94 | 2.06 | 2.91 | 4.03 | 2.64 | 1.63 | 4.09 | 0.00 | 0.83 | 4.46 |
| **pdb14** | 2.60 | 1.98 | 4.25 | 1.61 | 0.59 | 1.71 | 1.81 | 2.72 | 4.57 | 2.54 | 1.73 | 4.28 | 0.83 | 0.00 | 3.93 |
| **pdb15** | 2.97 | 4.18 | 3.63 | 3.32 | 4.16 | 3.56 | 3.36 | 3.91 | 3.39 | 4.57 | 2.02 | 3.46 | 4.46 | 3.85 | 0.00 |

## Supplementary scripts

Contains 6 example scripts:

```
setup_search.sh
start_runs_setup.sh
eval_result.sh
create_searchmodel_variations.sh
phaser_and_analysis_for_setup.sh
evaluate_for_setup.sh
```

*Brief guide:*

The scripts are hopefully relatively self-explanatory. The example scripts here show 3 datasets being tested with 4 scaffolds, each having 2 model variations (all atoms and poly-alanine), using 2 resolution-limits and 2 r.m.s.d. values on a 12 cpu-core cluster (a total of 240 runs). The scripts assumes the directory structure and file-placement mentioned below which should be created manually. Follow the steps noted here to initiate the MRPM search:

1) mkdir $MRPM (it is the root directory and can have any name).

2) mkdir $MRPM/models.

3) mkdir $MRPM/input.

4) Place 'setup_search.sh', 'start_runs_setup.sh', 'eval_result.sh' in $MRPM.

5) Place 'phaser_and_analysis_for_setup.sh' and 'evaluate_for_setup.sh' in $MRPM/input.

6) Place 'create_searchmodel_variations.sh' in $MRPM/models.

7) Place 'target.fas' (containing the target-sequence in FASTA) in $MRPM/input.

8) Place all datasets to test in $MRPM/input and name then data1.mtz, data2.mtz etc.

   Datasets should contain the following columns: H,K,L,FP,SIGFP.

10) Place the HA dataset to search for anomalous peaks in $MRPM/input and name ha-data.mtz.

   HA Dataset should contain the following columns: H,K,L,DANO,SIGDANO.

11) In $MRPM/models, create subdirectories called scaffold1, scaffold2 etc. One for each scaffold to test. Copy the pdb's to use as domains and scaffold into each scaffold-subdirectory.

12) Using pymol or similar, overlay the domains to the scaffold and save the final result as searchmodel.pdb.

12) Edit and run $MRPM/models/create_searchmodel.sh to generate the search-model library.

13) Edit $MRPM/input/phaser_and_analysis_for_setup.sh to set up the parameters to scan in the individual MR runs.

14) Edit $MRPM/input/evaluate_for_setup.sh to set up the parameters used to calculate the anomalous difference maps.

15) Edit and run $MRPM/setup_search.sh to set up the directory structure and input files for the search.

16) Edit and run $MRPM/start_runs_setup.sh to set up a 'start_runs.sh' file that will initiate MRPM on a given number of cores.

17) Run $MRPM/start_runs.sh to initiate MRPM.

18) During and after the runs have finished run $MRPM/eval_result.sh to list the results from individual runs that have completed. Use GNUPLOT or similar to plot the results if desired.

## Script 1: setup_search.sh

```
#!/bin/sh
#######################
#
# file 'setup_search.sh'
# setup the final data-structure and input files before initiating MRPM
# by Bjørn Panyella Pedersen,
# PUMPKIN centre, Aarhus University
#
#######################

dataarray="1 2 3"
scaffoldarray="1 2 3 4"
modelarray="1 2"

for d in ${dataarray}
do
    for s in ${scaffoldarray}
    do
        for m in ${modelarray}
        do
            # set up directory structure
            if [ ! -d ./data${d} ]; then mkdir data${d};fi
            cd data${d}
            if [ ! -d ./scaffold${s} ]; then mkdir scaffold${s};fi
            cd scaffold${s}
            if [ ! -d ./model${m} ]; then mkdir model${m};fi
            cd model${m}
            if [ ! -d ./output ]; then mkdir output;fi
            # now get the script
            cp ../../../input/phaser_and_analysis_for_setup.sh ./phaser_and_analysis.sh
            # set the dataset
            sed -i -e "s/<data>/data${d}\.mtz/" ./phaser_and_analysis.sh
            # set the scaffold
            sed -i -e "s/<scaffold>/scaffold${s}/" ./phaser_and_analysis.sh
            # set the model
            sed -i -e "s/<model>/model${m}\.pdb/" ./phaser_and_analysis.sh
            # Get the HA-peak evaluation script
            cp ../../../input/evaluate_for_setup.sh ./evaluate.sh
            # return to start
            cd ../../..
        done
    done
done
echo ""
echo " All done..."
echo ""
```

## Script 2: start_runs_setup.sh

```
#! /bin/sh
#######################
#
# file 'start_runs_setup.sh'
# setup to run multiple MRPM jobs on multiple cpus
# by Bjørn Panyella Pedersen,
# PUMPKIN centre, Aarhus University
#
#######################

noofcpu="12"
dataarray="1 2 3"
scaffoldarray="1 2 3 4"
modelarray="1 2"

echo ""
echo " Remember to edit this file to fit the experiment."
echo ""
echo " ___input___"
echo " noofcpu: $noofcpu"
echo " dataarray: $dataarray"
echo " scaffoldarray: $scaffoldarray"
echo " modelarray: $modelarray"

#cleanup
touch tmp.runlist start_runs.sh .cpu_tmp
rm tmp.runlist start_runs.sh .cpu_*

for d in ${dataarray}
do
    for s in ${scaffoldarray}
    do
        for m in ${modelarray}
        do
            echo "cd ./data${d}/scaffold${s}/model${m};sh phaser_and_analysis.sh;cd ../../.." >>tmp.runlist
        done
```

```
    done
done

# split up the runs
split -l $(echo $(cat tmp.runlist | wc -l)/${noofcpu} +1| bc) tmp.runlist .cpu_

for file in `echo ./.cpu_*`
do
echo "nohup sh $file >${file}_log &" >>start_runs.sh
done

#cleanup
rm tmp.runlist

echo ""
echo " run start_runs.sh to execute the search"
echo " All done..."
echo ""
```

### Script 3: eval_result.sh

```
#! /bin/sh
######################
#
# file 'eval_result.sh'
# Evaluate MRPM
# by Bjørn Panyella Pedersen,
# PUMPKIN centre, Aarhus University
#
######################
if [ ! $# = "1" ]; then
  echo ""
  echo " usage 'eval_result.sh <sort-keyword>'"
  echo " options: none name llg z ha"
  echo ""
  echo " use none for a quick view since no sort-argument is called"
  echo "only 'data*' directories are searched for solutions"
  echo ""
  exit
fi
# set key
key=$1

# get total number of runs
ta=`grep "array=" input/phaser_and_analysis_for_setup.sh | awk '{printf("*%s", NF)}' | xargs echo "1"|bc`
tb=`grep "array=" setup.sh | awk '{printf("*%s", NF)}' | xargs echo "1"|bc`
total=`echo "$ta*$tb"|bc`

# run the find command
if [ "${key}" = "none" ]; then
  find ./data* -name \*.summary | xargs cat >junktmp
  sol=`cat junktmp | wc -l`
  fail=`grep " -" junktmp | wc -l`
  part=`grep " yes" junktmp | wc -l`
  true=`echo "$sol-$fail-$part" |bc`
  cat junktmp
  rm junktmp
  echo " $sol/${total} runs completed ($true solutions, $part partial solutions and $fail with no solution)"
elif [ "${key}" = "name" ]; then
  find ./data* -name \*.summary | xargs cat >junktmp
  sort -n -k1 junktmp >junktmp2
  sol=`cat junktmp2 | wc -l`
  fail=`grep " -" junktmp | wc -l`
  part=`grep " yes" junktmp | wc -l`
  true=`echo "$sol-$fail-$part" |bc`
  cat junktmp2
  rm junktmp junktmp2
  echo " $sol/${total} runs completed ($true solutions, $part partial solutions and $fail with no solution)"
elif [ "${key}" = "llg" ]; then
  find ./data* -name \*.summary | xargs cat >junktmp
  sort -n -k14 junktmp >junktmp2
  sol=`cat junktmp2 | wc -l`
  fail=`grep " -" junktmp | wc -l`
  part=`grep " yes" junktmp | wc -l`
  true=`echo "$sol-$fail-$part" |bc`
  cat junktmp2
  rm junktmp junktmp2
  echo " $sol/${total} runs completed ($true solutions, $part partial solutions and $fail with no solution)"
elif [ "${key}" = "z" ]; then
  find ./data* -name \*.summary | xargs cat >junktmp
  sort -n -k16 junktmp >junktmp2
  sol=`cat junktmp2 | wc -l`
  fail=`grep " -" junktmp | wc -l`
  part=`grep " yes" junktmp | wc -l`
  true=`echo "$sol-$fail-$part" |bc`
  cat junktmp2
  rm junktmp junktmp2
```

```
    echo " $sol/${total} runs completed ($true solutions, $part partial solutions and $fail with no solution)"
  elif [ "${key}" = "ha" ]; then
    find ./data* -name \*.summary | xargs cat >junktmp
    sort -n -k18 junktmp >junktmp2
    sol=`cat junktmp2 | wc -l`
    fail=`grep " -" junktmp | wc -l`
    part=`grep " yes" junktmp | wc -l`
    true=`echo "$sol-$fail-$part" |bc`
    cat junktmp2
    rm junktmp junktmp2
    echo " $sol/${total} runs completed ($true solutions, $part partial solutions and $fail with no solution)"
else
  echo " sort-keyword unknown"
  echo " options: none name llg z ha"
fi
```

## Script 4: create_searchmodel_variations.sh

```
#!/bin/sh
#######################
#
# file 'create_searchmodel_variations.sh'
# merge and renumber scaffold input to generate models for MRPM
# by Bjørn Panyella Pedersen,
# PUMPKIN centre, Aarhus University
#
# model1: full model
# model2: full model, polyalanine
#
#######################

# loop though the following scaffolds:
i="1"
max="4"
while [ ${i} -le ${max} ]
do

# test that the scaffold dir exist
if [ ! -d ./scaffold${i} ]; then echo " FATAL ERROR: ./scaffold${i}/ is missing";exit; fi

cd ./scaffold${i}

# test for needed input
if [ ! -e searchmodel.pdb ]; then echo " FATAL ERROR: ./scaffold${i}/searchmodel.pdb is missing";exit; fi

# cleanup the input pdb
#remove AnisoU records from searchmodel and keep only chain A
pdbcur xyzin searchmodel.pdb xyzout searchmodel_tmp1.pdb <<EOF
lvchain A
noanisou
delhydrogen
delsolvent
end
EOF

# further cleanup
grep -v "CONECT" searchmodel_tmp1.pdb >searchmodel_tmp2.pdb
grep -v "MASTER" searchmodel_tmp2.pdb >searchmodel_tmp1.pdb

#reset bfactor
pdbset xyzin searchmodel_tmp1.pdb xyzout searchmodel_tmp2.pdb <<EOF
bfactor 50
end
EOF

# Model1, no changes
cp searchmodel_tmp2.pdb model1.pdb

# Model2, polyalanine version
pdbset XYZIN model1.pdb XYZOUT model2.pdb <<EOF
excl sidech
end
EOF

##
#
# insert additional search-model variations here as needed
#
##

# final cleanup
rm *tmp*

cd ..
echo ""
echo "  scaffold${i} done..."
echo ""
```

```
i=`echo "${i} + 1" | bc`
done

echo ""
echo "  All done..."
```

## Script 5: phaser_and_analysis_for_setup.sh

```
#! /bin/sh
#######################
#
# file 'phaser_and_analysis_for_setup.sh'
# input file to run the actual phaser-job (setup for phenix.phaser v2.3)
# by Bjørn Panyella Pedersen,
# PUMPKIN centre, Aarhus University
#
#######################

resarray="0.0 6.0 8.0"
rmsdarray="2 3"
pack="30"
number="1"
topfiles="10"
data="../../../input/<data>"
seq="../../../input/target.fas"
model="../../../models/<scaffold>/<model>"

# get dataid and pdbid for the final output.
dataid=`pwd | awk -F "/" '{printf("%s", $(NF-2))}'`
scaffoldid=`pwd | awk -F"/" '{printf("%s", $(NF-1))}'`
modelid=`pwd | awk -F"/" '{printf("%s", $NF)}'`
run="1"

for res in ${resarray}
do
    for rmsd in ${rmsdarray}
    do
        if [ ! -e ./result_run${run}.summary ]; then phenix.phaser << EOF > phaser_run${run}.log; fi
MODE MR_AUTO
HKLIN $data
LABIN  F=FP SIGF=SIGFP
TITLE find target using ${model}.pdb
COMPOSITION PROTEIN SEQ $seq &
    NUMBER $number
RESOLUTION 100.0 $res
ENSEMBLE $model &
    PDBFILE ${model}.pdb &
    RMS $rmsd
SEARCH ENSEMBLE $model NUMBER $number
PACK CUTOFF $pack
PACK SELECT ALLOW
FINAL ROT SELECT PERCENT 75.0
FINAL TRA SELECT PERCENT 75.0
SAVE ROT CLUSTER ON DUMP 20
SAVE TRA CLUSTER ON DUMP 20
PERMUTATIONS OFF
ROOT ./output/phaser_run${run}
HKLOUT ON
TOPFILES $topfiles
EOF

        # now calculate the HA-anom map..
        if [ ! -e ./result_run${run}.summary ]; then evaluate.sh ${run}; fi

        # now output the result in a nice simple way
        llg="-"
        z="-"
        ha="-"
        sol="-"
        z2="-"
        if [ -e output/phaser_run${run}.1.pdb ]; then llg=`awk 'NR==3 {printf("%s", $(NF-1))}' output/phaser_run$
{run}.1.pdb | cut -c5-`; fi
        if [ -e output/phaser_run${run}.1.pdb ]; then z=`awk 'NR==3 {printf("%s", $NF)}' output/phaser_run$
{run}.1.pdb | cut -c5-`; fi
        if [ -e result_run${run}.txt ]; then ha=`awk 'NR==2 {printf("%s", $2)}' result_run${run}.txt `; fi
        if [ -e result_run${run}.txt ]; then sol=`awk 'NR==2 {printf("%s", $1)}' result_run${run}.txt | cut -d"-"
-f2 | bc`; fi
        if [ -e result_run${run}.txt ]; then z2=`awk 'NR==3 {printf("%s", $3)}' output/phaser_run${run}.${sol}.pdb
| cut -c5-`; fi
        printf "data: %-6s  scaffold: %-10s  model: %6s  run: %2s  res: %3s  rmsd: %1s  LLG: %4s  Z: %4s  HA: %4s
from_sol: %2s with_z: %4s\n" $dataid $scaffoldid $modelid $run $res $rmsd $llg $z $ha $sol $z2 >result_run$
{run}.summary
        run=`echo "${run} + 1" | bc`
    done
done
```

**Script 6: evaluate_for_setup.sh**

```
#! /bin/sh
#######################
#
# file 'evaluate_for_setup.sh'
# file to run the HA anomalous map calculations
# by Bjørn Panyella Pedersen,
# PUMPKIN centre, Aarhus University
#
#######################

if [ $# = "0" ]; then
  echo ""
  echo "ERROR"
  echo ""
  echo "usage 'evaluate.sh <phaserrun-number to test> [purge]'"
  echo "purge keywork is optional and will force rewrite of all data"
  echo "output is the top anomalous peak from each phaser solution"
  echo ""
  exit
fi

# HA data to test against
data="../../../input/ha-data.mtz"

# fft ano map cutoff to test (3 currently allowed)
cutoff1="9"
cutoff2="7.5"
cutoff3="6"

# number of max solutions from phaser to test
maxsoltotest="10"

# set run number
run=$1

# ugly fix
touch ./output/phaser_run${run}.1.mtz

# start analysis
for model in `echo ./output/phaser_run${run}.*.mtz`
do

# exit if there was no solution
if [ ! -s ./output/phaser_run${run}.1.mtz ];then rm ./output/phaser_run${run}.1.mtz; exit; fi

# get pdb name
pdb=`basename "$model" .mtz`
# get solution number
sol=`basename "$model" | cut -d "." -f2`
# get FOM
fom=`mtzdmp $model | grep "FOM" | tail -n1 | awk '{printf($7)}'`

# cad phaser-phases to dataset with anomalous data
if [ ! -e ./output/cad_run${run}_sol${sol}.mtz ]; then cad HKLIN1 $model HKLIN2 $data HKLOUT ./output/cad_run$
{run}_sol${sol}.mtz <<EOF >./output/cad_run${run}_sol${sol}.log; fi

LABIN FILE 1 E1 = PHIC E2 = FOM
LABOUT FILE 1 E1 = PHIC E2 = FOM
CTYPEIN FILE 1 E1 = P E2 = W
LABIN FILE 2 E1 = DANO E2 = SIGDANO
LABOUT FILE 2 E1 = DANO E2 = SIGDANO
CTYPEIN FILE 2 E1 = D E2 = Q
END
EOF

#fft to get anomalous difference peaks

############
## cutoff1
if [ ! -e ./output/fft_run${run}_sol${sol}_cutoff${cutoff1}.log ]; then fft HKLIN ./output/cad_run${run}_sol$
{sol}.mtz MAPOUT ./output/fft_run${run}_sol${sol}_cutoff${cutoff1}.map <<EOF >./output/fft_run${run}_sol$
{sol}_cutoff${cutoff1}.log; fi
XYZLIM ASU
SCALE F1 1.0
#SCALE F2 1.0
RESOLUTION 100 $cutoff1
LABIN DANO=DANO SIG1 = SIGDANO PHI=PHIC W=FOM
END
EOF

#move map to model
if [ ! -e ./output/mapmask_run${run}_sol${sol}_cutoff${cutoff1}.log ]; then mapmask MAPIN ./output/fft_run$
{run}_sol${sol}_cutoff${cutoff1}.map MAPOUT ./output/mapmask_run${run}_sol${sol}_cutoff${cutoff1}.map XYZIN
./output/${pdb}.pdb <<EOF >./output/mapmask_run${run}_sol${sol}_cutoff${cutoff1}.log; fi
BORDER 20
END
```

```
EOF

#peakmax
if [ ! -e ./output/peakmax_run${run}_sol${sol}_cutoff${cutoff1}.log ]; then peakmax MAPIN ./output/mapmask_run$
{run}_sol${sol}_cutoff${cutoff1}.map XYZOUT ./output/peakmax_run${run}_sol${sol}_cutoff${cutoff1}.pdb XYZFRC
./output/peakmax_run${run}_sol${sol}_cutoff${cutoff1}.ha <<EOF >./output/peakmax_run${run}_sol${sol}_cutoff$
{cutoff1}.log; fi
THRESHOLD RMS 1.0
NUMPEAKS 50
OUTPUT BROOKHAVEN FRAC
RESIDUE WAT
ATNAME OW
CHAIN X
END
EOF

# grep for first peak
peak=""
peak=`grep "ATOM1 " ./output/peakmax_run${run}_sol${sol}_cutoff${cutoff1}.ha | awk '{print $6}'`

if [ "$peak" = "" ]; then
  peak="n/a"
fi

#print result
printf "%3s-%.2d %5s %6s %11s\n" $run ${sol} $peak $fom $cutoff1 >>tmp3${run}.txt

############
## cutoff2
if [ ! -e ./output/fft_run${run}_sol${sol}_cutoff${cutoff2}.log ]; then fft HKLIN ./output/cad_run${run}_sol$
{sol}.mtz MAPOUT ./output/fft_run${run}_sol${sol}_cutoff${cutoff2}.map <<EOF >./output/fft_run${run}_sol$
{sol}_cutoff${cutoff2}.log; fi
XYZLIM ASU
SCALE F1 1.0
#SCALE F2 1.0
RESOLUTION 100 $cutoff2
LABIN DANO=DANO SIG1 = SIGDANO PHI=PHIC W=FOM
END
EOF

#move map to model
if [ ! -e ./output/mapmask_run${run}_sol${sol}_cutoff${cutoff2}.log ]; then mapmask MAPIN ./output/fft_run$
{run}_sol${sol}_cutoff${cutoff2}.map MAPOUT ./output/mapmask_run${run}_sol${sol}_cutoff${cutoff2}.map XYZIN
./output/${pdb}.pdb <<EOF >./output/mapmask_run${run}_sol${sol}_cutoff${cutoff2}.log; fi
BORDER 20
END
EOF

#peakmax
if [ ! -e ./output/peakmax_run${run}_sol${sol}_cutoff${cutoff2}.log ]; then peakmax MAPIN ./output/mapmask_run$
{run}_sol${sol}_cutoff${cutoff2}.map XYZOUT ./output/peakmax_run${run}_sol${sol}_cutoff${cutoff2}.pdb XYZFRC
./output/peakmax_run${run}_sol${sol}_cutoff${cutoff2}.ha <<EOF >./output/peakmax_run${run}_sol${sol}_cutoff$
{cutoff2}.log; fi
THRESHOLD RMS 1.0
NUMPEAKS 50
OUTPUT BROOKHAVEN FRAC
RESIDUE WAT
ATNAME OW
CHAIN X
END
EOF

# grep for first peak
peak=""
peak=`grep "ATOM1 " ./output/peakmax_run${run}_sol${sol}_cutoff${cutoff2}.ha | awk '{print $6}'`

if [ "$peak" = "" ]; then
  peak="n/a"
fi

#print result
printf "%3s-%.2d %5s %6s %11s\n" $run ${sol} $peak $fom $cutoff2 >>tmp3${run}.txt

############
## cutoff3
if [ ! -e ./output/fft_run${run}_sol${sol}_cutoff${cutoff3}.log ]; then fft HKLIN ./output/cad_run${run}_sol$
{sol}.mtz MAPOUT ./output/fft_run${run}_sol${sol}_cutoff${cutoff3}.map <<EOF >./output/fft_run${run}_sol$
{sol}_cutoff${cutoff3}.log; fi
XYZLIM ASU
SCALE F1 1.0
#SCALE F2 1.0
RESOLUTION 100 $cutoff3
LABIN DANO=DANO SIG1 = SIGDANO PHI=PHIC W=FOM
END
EOF

#move map to model
if [ ! -e ./output/mapmask_run${run}_sol${sol}_cutoff${cutoff3}.log ]; then mapmask MAPIN ./output/fft_run$
```

```
{run}_sol${sol}_cutoff${cutoff3}.map MAPOUT ./output/mapmask_run${run}_sol${sol}_cutoff${cutoff3}.map XYZIN
./output/${pdb}.pdb <<EOF >./output/mapmask_run${run}_sol${sol}_cutoff${cutoff3}.log; fi
BORDER 20
END
EOF

#peakmax
if [ ! -e ./output/peakmax_run${run}_sol${sol}_cutoff${cutoff3}.log ]; then peakmax MAPIN ./output/mapmask_run$
{run}_sol${sol}_cutoff${cutoff3}.map XYZOUT ./output/peakmax_run${run}_sol${sol}_cutoff${cutoff3}.pdb XYZFRC
./output/peakmax_run${run}_sol${sol}_cutoff${cutoff3}.ha <<EOF >./output/peakmax_run${run}_sol${sol}_cutoff$
{cutoff3}.log; fi
THRESHOLD RMS 1.0
NUMPEAKS 50
OUTPUT BROOKHAVEN FRAC
RESIDUE WAT
ATNAME OW
CHAIN X
END
EOF

# grep for first peak
peak=""
peak=`grep "ATOM1 " ./output/peakmax_run${run}_sol${sol}_cutoff${cutoff3}.ha | awk '{print $6}'`

if [ "$peak" = "" ]; then
  peak="n/a"
fi

#print result
printf "%3s-%.2d %5s %6s %11s\n" $run ${sol} $peak $fom $cutoff3 >>tmp3${run}.txt

# sort the 3 cutoff lines and only use the best one.
sort -k2 -r tmp3${run}.txt >>result_sorted1_run${run}.txt
head -n 1 result_sorted1_run${run}.txt >> tmp${run}.txt
rm tmp3${run}.txt result_sorted1_run${run}.txt

done

# cleanup
rm output/*_run${run}_*

# process final result
sort -k2 -r tmp${run}.txt >>tmp2_run${run}.txt
echo "   run  peak     FOM  fft-cutoff" >result_run${run}.txt
cat tmp2_run${run}.txt >>result_run${run}.txt
rm tmp${run}.txt tmp2_run${run}.txt
```